\documentclass[useAMS,usenatbib]{mnras}

\usepackage{bm}
\usepackage{natbib}
\usepackage{graphicx}
\usepackage{amssymb, amsmath}
\usepackage[normalem]{ulem}
\usepackage{xcolor}

\title[Deep crustal heating for realistic 
compositions
of thermonuclear ashes]{Deep crustal heating for realistic 
compositions
of thermonuclear ashes}
\author[N.N.\ Shchechilin, M.E.\ Gusakov, A.I.\ Chugunov]
{N.N.\ Shchechilin,\thanks{nicknicklas@mail.ru} M.E.\ Gusakov, A.I.\ Chugunov\\
Ioffe Institute, Polytekhnicheskaya 26, 194021 Saint Petersburg, Russia}

\begin{document}
	
	\date{Accepted 2021 xxxx. Received 2021 xxxx;
		in original form 2021 xxxx}
	
	\pagerange{\pageref{firstpage}--\pageref{lastpage}}
	\pubyear{2021}
	
	\maketitle
	
	\label{firstpage}

\begin{abstract}
The deep crustal heating, 
associated with exothermal nuclear reactions, 
is believed to be a key parameter
for describing the thermal evolution 
of accreting neutron stars. 
In this paper, 
we present the first 
thermodynamically consistent
calculations of the crustal heating 
for realistic compositions of thermonuclear ashes. 
In contrast to previous studies 
based on the traditional approach, 
we account for neutron hydrostatic/diffusion (nHD) equilibrium condition
imposed by superfluidity of neutrons in a major part of the inner crust and rapid diffusion in the 
remaining part of the inner crust.
We apply a simplified reaction network to model nuclear evolution of various multi-component thermonuclear burning ashes (superburst, KEPLER, and extreme rp-process ashes) in the outer crust and calculate the deep crustal heating energy release $Q$, parametrized by the pressure at the outer-inner crust interface, $P_\mathrm{oi}$. 
Using the general thermodynamic arguments, we set a lower limit on $Q$, 
$Q\gtrsim 0.13-0.2$~MeV per baryon 
(an actual value depends on the ash composition and the employed mass model).
\end{abstract}

\begin{keywords}
	stars: neutron, accretion discs, X-rays: binaries
\end{keywords}

\section{Introduction}

In neutron star (NS) low-mass X-ray binaries (LMXBs) a mass transfer from a less compact donor star (typically, with the mass $\lesssim M_\odot$) via Roche-lobe overflow triggers numerous observable phenomena. One of them is the thermonuclear burning of light elements on the NS surface. Depending on accreted matter composition and accretion rate, burning proceeds in different regimes (\citealt{johnston20}) and produces ashes, consisting of large number of heavy
nuclei
(see, e.g., \citealt{Schatz_ea01,KH11,Cyburt_ea16}). Afterwards, the newly-fallen material pushes the ashes 
inward the star,
so that their compression induces exothermic nuclear reactions (\citealt{Sato79,HZ90,GC20_DiffEq}). 
These reactions heat up the NS crust, 
leading to the observable thermal emission in quiescent periods between the accretion episodes (\citealt*{bbr98,lh07}; \citealt{heinke_et_al_09}; \citealt*{pcc19}; \citealt{Parikh_ea21,Parikh_ea21_erratum}). 

Comparison of the NS crust cooling curves to the theoretical models of heat relaxation 
(\citealt{Ruthledge_etal02_KS,syhp07,bc09,Cackett_ea10,pr13}; \citealt*{degenaar_ea13}; \citealt{mdkse18,Parikh_ea19,Wijngaarden_ea20,pc21}),
as well as modelling of the 
steady-state thermal configurations 
(\citealt{bbr98}; \citealt*{ylh03}; \citealt{ylpgc04,hjwt07,heinke_et_al_09}; \citealt*{wdp13}; \citealt{by15,hs17,Fortin_ea18,Brown_ea18,pcc19,Fortin_ea21})
allows one to constrain the properties of NS matter. 
To model the thermal evolution one needs to know the profile of energy release, 
the crust equation of state (EOS) and composition, which can be 
found by studying the chain of nuclear reactions in the crust of an accreting NS. 

First calculations of the accreted crust structure were performed
by \cite{Sato79}. Later, \cite{HZ90,HZ90b,HZ03,HZ08} constructed a set of the most widely-used models, based on the compressible liquid-drop model (CLD) of \cite{MB77}. These models were recently updated by \cite{Fantina_ea18}, who implemented realistic Skyrme-type parametrization of nucleon interactions suggested by \cite{Goriely_ea10} 
and work within the semiclassical extended Thomas-Fermi approach (see, e.g., \citealt{Brack_ea85}) with shell corrections included `on top' by \cite{Str67} integral theorem. These studies adopted a one-component approximation, assuming that nuclei of only one type exist at any given pressure. \cite{Gupta_ea07,gkm08} and \cite{Steiner12} went beyond the one-component
approximation. Recently, \cite{lau_ea18} presented detailed calculations of the crust composition and heat release down to the depth in the inner crust with the density $\rho\lesssim2\times10^{12}$\,g cm$^{-3}$. 
They used full reaction network 
and realistic compositions of nuclear ashes.
Finally, \cite{SC19_MNRAS} investigated the dependence of the crust parameters on the employed mass model using the simplified reaction network and found a
reasonable
agreement with the results of \cite{lau_ea18} for initial $^{56}$Fe composition, if the same mass model is applied.

All the papers quoted above consider nuclear reactions in 
a compressing matter element, keeping the number of baryons in this element fixed during the compression (the traditional approach). Inconsistency of the traditional approach was pointed out by \cite{CS19_NoEquil}: it neglects possible redistribution of unbound neutrons in the inner crust. The way out was suggested by \cite{GC20_DiffEq}, who developed a thermodynamically consistent approach, that accounts for redistribution of neutrons. The approach rests on the neutron Hydrostatic and Diffusion (nHD) equilibrium condition in the inner crust,
$\mu_\mathrm{n}^\infty=\mu_n e^{\nu/2}=\mathrm{const}$, where $\mu_\mathrm{n}$ is the (local) neutron chemical potential, and $e^{\nu/2}$ is the redshift factor. In the major part of the inner crust this condition should hold in order for the superfluid neutrons there to be in hydrostatic equilibrium; in turn, in the remaining nonsuperfluid region the condition is satisfied due to the efficient neutron diffusion (\citealt{GC20_DiffEq}).

Modelling 
of 
the accreted crust within the nHD approach appears to be much more intricate than in the traditional approach,
because 
one should self-consistently 
study
nuclear reactions in the whole inner crust, 
accounting for, simultaneously, 
the nHD and general hydrostatic equilibrium conditions.
To do so
one should 
treat the pressure $P_\mathrm{oi}$ at the outer-inner (oi) crust interface as a parameter, which, generally, does not coincide with the neutron-drip pressure, but 
varies
during the accretion process 
as EOS itself,
until the fully accreted (FA) crust is established
(\citealt{GC20_DiffEq}). For FA crust, the nuclear reactions proceed in the regime, which keeps the crust EOS without noticeable changes. In particular, the total number of nuclei in the crust is conserved (up to a small secular 
variation
associated with the 
increasing
NS mass). 

To keep the number of nuclei in the crust fixed, an efficient mechanism of nuclei disintegration is required. Such mechanism indeed exists in the form of a specific instability, which typically occurs in the bottom layers of the inner crust (see \citealt{GC20_DiffEq} for details). As discussed in \cite{GC21_HeatReleaze} 
(GC21 in what follows), the instability is affected by 
the nuclear
shell effects. Furthermore, part of the crust can be located below the instability, being decoupled from the rest of the crust (i.e., 
newly
accreted nuclei do not reach this region). The composition of this part of the crust is determined 
during accretion
 before the crust becomes fully accreted. Because of these difficulties, accurate prediction of the FA crust EOS within the thermodynamically consistent approach stays an open problem.

However,
as shown in GC21, the details of nuclear evolution 
in the inner crust are not so important, 
if one is only interested in the total deep crustal heating energy release $Q$ 
(see section \ref{Sec:Qi} for the accurate definition, 
which takes into account the redshift factors). Namely,  $Q$ can be calculated by modelling 
the outer crust EOS at $P<P_\mathrm{oi}$, where pressure $P_\mathrm{oi}$ is treated as a 
free
parameter, 
which encodes all information on the reactions in the inner crust, 
including the nuclei disintegration process (GC21). 
This allows one to 
separate 
the problem of $Q$
determination
into two steps: 
in the first step $Q$ is calculated as a function of $P_\mathrm{oi}$ and in the second step the value of $P_\mathrm{oi}$ is determined by accurate modelling of the crust evolution in the course of accretion. 
In this paper, we consider only the first step, leaving the second one for the subsequent work.

The paper is organized as follows. 
In section \ref{outcr} we describe 
the simplified reaction network applied 
to studying the
nuclear evolution in the outer crust. 
In section \ref{Sec:outercrust} we present results of this modelling for various initial compositions of the ashes and nuclear mass models, in particular, we calculate the heat release in the outer crust, $Q_\mathrm{o}$. 
In section \ref{Sec:Qi} we calculate the heat release below the outer crust, $Q_\mathrm{inner}$, 
and the net deep crustal heating energy release, $Q$. 
The quantities $Q_\mathrm{o}$,  $Q_\mathrm{inner}$  and $Q$ are  parametrized by the pressure $P_\mathrm{oi}$.
Additionally, in Appendix \ref{app_compos} we present profiles of the average charge and impurity parameter in the outer crust for all the considered models.

\section{Simplified reaction network 
in the
outer crust}\label{outcr}

In the outer crust, the pressure is almost completely determined by the degenerate relativistic electrons, i.e., the pressure growth is associated with the increasing electron chemical potential $\mu_\mathrm{e}$. Increase of $\mu_\mathrm{e}$ induces electron captures by nuclei (typically, the captures occur 
in a pairwise fashion
due to even-odd staggering, see, e.g., \citealt{HZ90}). 
For the considered compositions 
of nuclear ashes some nuclei become very neutron-rich at the bottom layers of the outer crust, so that they emit neutrons after electron captures. However, the total amount of the emitted neutrons is very small. Because neutron diffusion timescale is typically much larger than 
the characteristic time of neutron absorptions, we assume, that neutrons are captured by other nuclei, located at the same depth (a probability to adsorb neutron during the neutron-nuclear scattering is not small enough to allow for a large number of scattering events, required for neutron diffusion over any reasonable distance).
Another consequence of electron captures is the appearance of elements with a low proton number $Z$, 
for which pycnonuclear 
fusion 
reactions appear to be possible (\citealt{Yakovlev_ea06}).
In some cases, in a complex chain of reactions, the emission of electrons turns out to be energetically favourable.
Consequently, in the outer crust one should consider electron emissions/captures, neutron emissions/captures, and pycnonuclear 
fusions (for simplicity, in this work we neglect neutron transfer reactions, considered by  \citealt{SA72_n_transf,Chugunov19_n_transf}).

\subsection{Reaction network} \label{network}

Let us briefly remind the main points of the simplified reaction network discussed in \cite{SC19_MNRAS} and applied here. The approach, generally, follows that of \cite{Steiner12}, 
and is based on the minimization of the Gibbs energy at constant pressure, neglecting thermal corrections. We start with some initial pressure $P_\mathrm{in}$ and gradually increase it, step by step, mimicking compression of a given matter element in the outer crust. At each compression step,
we check for available reactions that can decrease the Gibbs energy. If some reaction is open, we accurately adjust the pressure to the reaction's threshold value. In this way, we avoid unphysical energy release associated with the stepwise increase of the pressure. After the first 
reaction occurred, the subsequent admissible reactions (if any) proceed at constant pressure by small chunks  until none of the reactions are energetically favourable. 
Then we decide that the matter composition at this $P$ is established 
and we can further increase the pressure.

In our simplified reaction network the order of the reaction chunks is controlled by the 
following	
priority rules:
(a) neutron emission, 
(b) neutron capture, 
(c) electron emission/capture plus (optional) neutron emission, 
(d) neutron capture and electron emission/capture, 
(e) capture of 2 neutrons, 
(f) pycnonuclear fusion.
Among the allowed reactions of the same type, 
the most energetically favourable reaction goes first. 
These priority rules are based on the estimates of the reaction rates (\citealt{SC19_MNRAS}) with few modifications. First, following \cite{Fantina_ea18} we now allow for the electron emissions/captures accompanied by neutron emissions/captures [reaction types (c) and (d)]. 
Second, the priority of the two-neutron-capture reaction is reduced.
As discussed in \cite{SC19_MNRAS}, this reaction was included into the simplified reaction network to mimic two successive neutron captures in the inner crust, if the first one is not energetically favourable, 
but proceeds in the (more realistic) detailed network due to thermal activation. 
Here we apply our network  
to
the outer crust,
where a very small amount of unbound neutrons is expected (see, e.g., figure 7 in \citealt{lau_ea18})
and it seems unnecessary to consider two-neutron-captures,
but the analysis of the reaction pathways shown in \cite{lau_ea18} 
reveals that this process takes place in their network. 
To include this possibility, 
we retain the reaction 
type (e) in our network, but with reduced priority. In numerical simulations the reaction (e) allows to absorb unbound neutrons, 
released by reactions
(a) and (c), if one-neutron-capture reaction is 
forbidden.
 
In our model the fusion reactions are only allowed if the respective timescale does not exceed the accretion time (i.e., the replacement time $\tau\approx P/(g \dot{m}_\mathrm{fid})$ for an accreted layer with the pressure $P$, where the fiducial accretion rate is taken to be $ \dot{m}_\mathrm{fid}=0.3\dot{m}_\mathrm{Edd}$; $\dot{m}_\mathrm{Edd}=8.8\times 10^4$\,g cm$^{-2}$ s$^{-1}$ is the Eddington accretion rate, and $g=1.85\times 10^{14}$\,cm s$^{-2}$ is the gravitational acceleration). To estimate the 
reaction timescale
we apply 
the thermally enhanced pycnonuclear reaction rates from \cite{Yakovlev_ea06} with $S$-factors from \cite{Afanasjev_ea12}. As in \cite{lau_ea18}, the temperature is assumed to be fixed, $T=5\times 10^8$\,K. Our reaction network depends on the temperature only via the fusion reaction rates.

Following \cite{HZ08} and \cite{Fantina_ea18}, we completely neglect neutrino losses when calculating the energy release for all nuclear reactions 
(this is justified by
the work of \cite{Gupta_ea07}, 
where excited 
states 
are taken into account). 
In \cite{SC19_MNRAS} we demonstrate 
that the simplified reaction network is
in a reasonable agreement with the results of detailed 
network 
of \cite{lau_ea18} for pure $^{56}$Fe composition of nuclear ashes (see also Fig.\ \ref{Fig_Lau} in Appendix \ref{app_compos} and section 3.3 in  \citealt{Gupta_ea07} for a discussion of applicability of approximate models).

\subsection{Physics input} \label{Sec:PhysInput}

To implement the simplified reaction network, one should specify the model 
for 
calculating the Gibbs energy 
for
a given nuclear composition. 
The physics of plasma in the outer crust of accreting neutron stars is rather well-known (see, e.g., \citealt{hpy07}; here we apply the same model as in \citealt{SC19_MNRAS}) so that the masses of nuclei become the main source of uncertainty. If available, we use experimental results from the Atomic Mass Evaluation 2020 (AME20; \citealt{ame20}).%
\footnote{The table for AME20 is downloaded from \url{https://www-nds.iaea.org/amdc/}.
Extrapolated mass values from this table are ignored.
} 
For comparison,
the
results based on the previous version of the atomic mass evaluation (AME16, \citealt{ame16}) are shown in Figs.\
\ref{Fig_AME16_Q} and \ref{Fig_AME16_mu} in Appendix \ref{app_massModels}, and also discussed in depth in the first arXiv version of this paper.\footnote{\url{https://arxiv.org/pdf/2105.01991v1.pdf}}

However, in the bottom regions of the outer crust we also need to use the theoretical mass models. 
Despite the fact that the modern mass models 
agree well with the experimental data (rms deviation is about $0.6$ MeV), the mass difference between particular elements can reach $\sim 10$~MeV
(see Fig.\ \ref{Fig_MB} in Appendix \ref{app_massModels}). 
To check the degree of model independence 
of our results, we apply several theoretical atomic mass tables, namely, two finite-range droplet models, FRDM92  \cite{FRDM95} and FRDM12 \cite{FRDM12}, together with the Hartree-Fock-Bogoliubov model HFB24 from \cite{Goriely_ea_Bsk22-26}.%
\footnote{The tables for FRDM92, FRDM12 and HFB24 models are downloaded from \url{http://t2.lanl.gov/nis/molleretal/publications/ADNDT-59-1995-185-files.html},
\url{http://t2.lanl.gov/nis/molleretal/publications/ADNDT-FRDM2012.html},	
and \url{http://www.astro.ulb.ac.be/bruslib/nucdata/hfb24-dat} (\citealt{bruslib}), respectively.}
To obtain the masses of nuclei, the electron rest mass and binding energy $E_\mathrm{e,b}$ is subtracted using the fitting formula $E_\mathrm{e,b}=14.33 Z^{2.39}$\,eV (see \citealt{Lunney_ea03_elbind}) for all mass tables. 
In our calculations, we 
use
the mass tables obtained by 
combining
AME20 with the respective theoretical mass tables
(the `joint' approach;  
see
\citealt{SC19_MNRAS} 
for details and discussion of mass table merging).
For simplicity, we include only ground state to ground state transitions, that allow us to avoid uncertainty related to the structure of excited  
states.

\subsection{Initial composition and numerical implementation} 
\label{Sec:num}   

The thermonuclear burning in the shallow 
crust regions can occur in different regimes, 
leading to various ash compositions (\citealt{johnston20}). 
Following \cite{lau_ea18}, 
we consider three representative models of nuclear ashes: 
extreme rapid proton capture (extreme rp-process) ashes (\citealt{Schatz_ea01}), KEPLER X-ray burst ashes (\citealt{Cyburt_ea16}), and superburst ashes (\citealt{Cumming_ea06}). 
The corresponding ash compositions 
are extracted from figures 9, 15, 18 of \cite{lau_ea18}.
Note that, the most abundant elements in the superburst and KEPLER ashes are from the same group of elements with the mass number $A\approx60$ ($\rm{^{56}}Fe$ and $\rm{^{64}}Ni$ for superburst and KEPLER ashes, respectively). In turn, for extreme rp-process ashes the abundance peak is shifted to $A=106$ 
(and is dominated by $\rm{^{106}}Ru$), due to efficient generation of heavy elements by rapid proton captures. 
 
For numerical simulations we take 500 nuclei, thereby the abundances from \cite{lau_ea18} are rounded; we checked that this choice does not affect our conclusions.%
%
\footnote{We use two independent codes, with a bit different realization of the minimization procedure. Both codes give similar results.} 
%
The charge number $Z$ for each $A$ in the ashes is chosen to be the most bound nuclei 
in the 
terrestrial conditions ($P=0$).
Then, rising the pressure to $P_\mathrm{in}=6\times10^{26}$\,dyn cm$^{-2}$ 
we obtain the initial set of nuclei to start our simulations.
The reaction network can 
be used
as described
in Sec.\ \ref{network}
up to the outer-inner crust interface, 
corresponding to the pressure $P_\mathrm{oi}$ and should be modified in the inner crust to take into account the nHD condition.

As shown in GC21, the pressure $P_\mathrm{oi}$ depends on the (still) rather uncertain
nuclear shell effects in the 
deep 
layers of the inner crust; 
it can also depend on the accretion history. Thus, here we treat $P_\mathrm{oi}$ as a free
parameter of the outer crust models, leaving construction of the inner crust models and accurate determination of $P_\mathrm{oi}$ 
as 
a task for the future.

Note that there is no need for recalculating 
the outer crust model for each given pressure $P_{\rm oi}$. Instead, it is instructive to proceed as follows. 
As a first step, we generate the traditional model of the outer crust 
by running our simplified reaction network up to the neutron drip pressure
$P^\mathrm{(acc)}_\mathrm{nd}$,
which is 
rigorously defined in the traditional approach 
(it
is the minimum pressure at which unbound neutrons appear in the accreted crust and are not captured by nuclei,
assuming that the nHD condition is ignored; \citealt{HZ90,HZ03,HZ08,Steiner12,lau_ea18,SC19_MNRAS}).
The obtained {\it traditional} outer-crust model 
does not have a direct physical meaning
as long as $P_{\rm oi}<P^\mathrm{(acc)}_\mathrm{nd}$,
but it will be helpful in the second step.
	Namely, at the second step, 
we specify $P_{\rm oi}$ and produce the realistic outer-crust model 
by cutting off the $P>P_\mathrm{oi}$ part of the 
traditional model generated at the first step. 
 
By construction, this two-step procedure is applicable 
if $P_\mathrm{oi}<P^\mathrm{(acc)}_\mathrm{nd}$; 
the latter condition agrees with the numerical results of 
(\citealt{GC20_DiffEq}, GC21),
which are obtained assuming pure $^{56}$Fe
ash composition. 
However, we should warn the reader, that we are not aware of a rigorous proof 
that $P_\mathrm{oi}$ can never exceed $P^\mathrm{(acc)}_\mathrm{nd}$. 
If, for some specific conditions (e.g., ash composition and/or inner-crust nuclear-physics model), $P_\mathrm{oi}$ will appear to be larger than $P^\mathrm{(acc)}_\mathrm{nd}$, 
the outer crust model at $P>P^\mathrm{(acc)}_\mathrm{nd}$
should be supplemented to account for
{\it sedimentation} of unbound neutrons into the inner crust.
We leave 
a detailed analysis of this (rather exotic) possibility
beyond the scope of the present work.

\section{Results}
\label{Sec:res}

\subsection{Nuclear evolution and heat release in the outer crust} \label{Sec:outercrust}


\begin{figure}
	\includegraphics[width=\columnwidth]{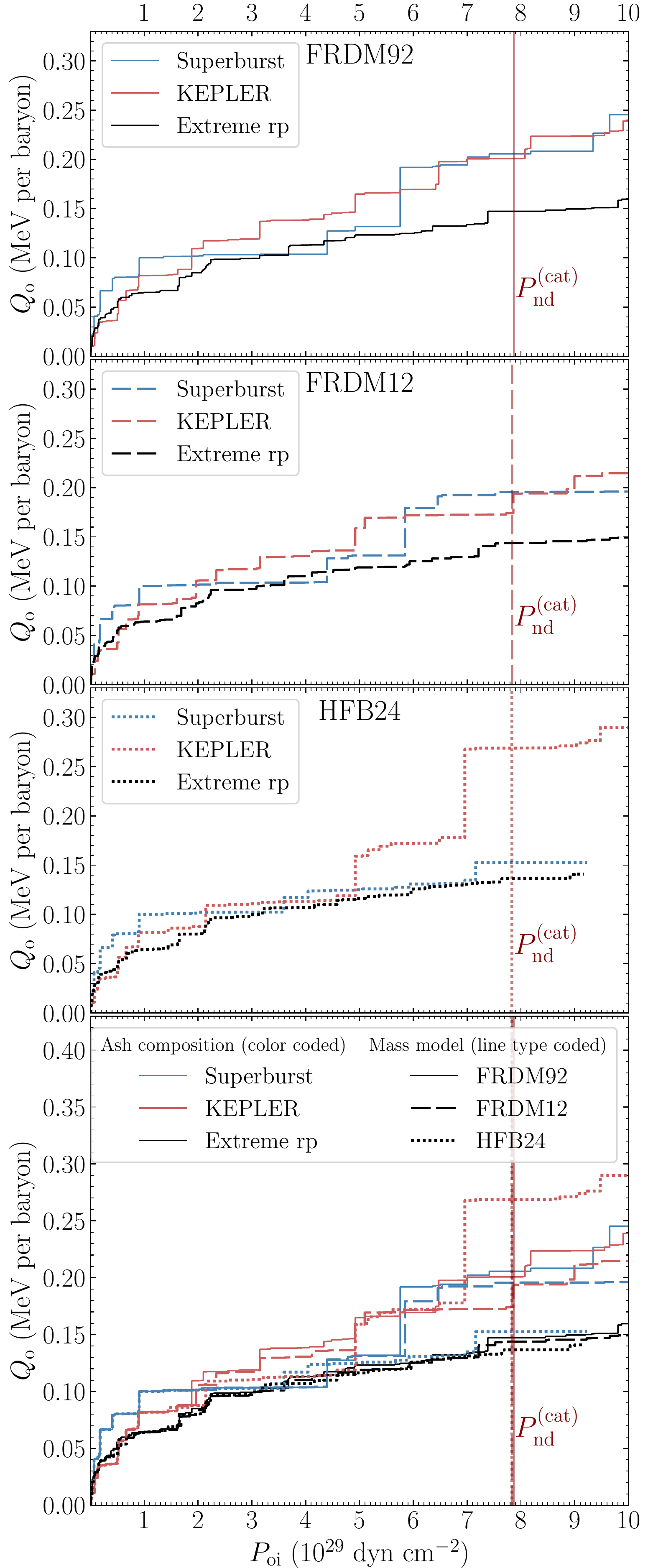}
	\caption{$Q_\mathrm{o}$ vs $P_\mathrm{oi}$
	for the three ash compositions considered in the paper. Three upper panels correspond to the three mass models (from the top: FRDM92, FRDM12, and HFB24), the bottom panel shows all the plots in one figure. Ash composition is colour coded: superburst -- blue; KEPLER -- red; extreme rp -- black; mass model is line type coded: FRDM92 -- solid; FRDM12 -- dashed; HFB24 -- dotted. The vertical lines represent the pressure $P_\mathrm{nd}^\mathrm{(cat)}$ at the oi interface for the cold catalysed matter (line type corresponds to the mass model).}
	\label{Fig:Qo}
\end{figure}

We calculate the nuclear composition 
in the outer 
crust by
applying AME20
for nuclei with experimentally known masses and one of the theoretical nuclear mass models (FRDM92, FRDM12, and HFB24) otherwise (see section \ref{Sec:PhysInput}). 
In Fig.\ \ref{Fig:Qo} we show
the heat release in the outer crust, $Q_\mathrm{o}(P_\mathrm{oi})$. 
For simplicity, as in all the previous works 
(e.g., \citealt{,HZ08,lau_ea18,Fantina_ea18}), 
we neglect variation of the redshift factor 
in the outer crust and calculate $Q_\mathrm{o}(P_\mathrm{oi})$ 
as a (simple) sum over the heat sources located 
at pressures lower than $P_\mathrm{oi}$.
Two curves
for HFB24 mass model terminate
at $P_\mathrm{oi}=P^\mathrm{(acc)}_\mathrm{nd}$ (for most of the models
$P^\mathrm{(acc)}_\mathrm{nd}$ is larger than $10^{30}$~dyn\,cm$^{-2}$, 
the maximum pressure shown in Fig.\ \ref{Fig:Qo}).

It is worth stressing, that the heat release profile for $P<P_\mathrm{oi}$ is not 
affected
by the nHD condition. 
As a result, the integrated heat release up to a given pressure $P$ can be easily read out from Fig.\ \ref{Fig:Qo} as $Q_\mathrm{o}(P)$.
Conversely, 
it allows one to map previously calculated heat release profiles (e.g., \citealt{HZ90,HZ03,HZ08,Steiner12,lau_ea18,Fantina_ea18,SC19_MNRAS,cfzh20}) to obtain $Q_\mathrm{o}(P_\mathrm{oi})$ dependence for these models.

Up to the pressure $\lesssim 5\times 10^{28}$ dyn cm$^{-2}$ the crust is mostly composed of nuclei with experimentally measured masses 
and the nuclear reaction chain is almost 
independent 
of the 
theoretical mass 
model
(see bottom panel in Fig.\ \ref{Fig:Qo}).
To extend 
the calculation
to higher pressures 
theoretical mass tables are required and, as a consequence, $Q_\mathrm{o}(P_\mathrm{oi})$ 
curves
start branching, indicating sensitivity 
of the results
to the 
employed
mass model. 
For instance, for superburst ashes $Q_\mathrm{o}$ predicted by HFB24 and FRDM92 mass models can differ by a factor of 1.3 for
$P_\mathrm{oi}\approx 8\times10^{29}$~dyn\,cm$^{-2}$ 
(about $0.05$~MeV
per accreted baryon).
This difference is mainly associated with the model-dependent $Q$-values of nuclear reactions. 
In this case, the predominant contribution comes from the reaction $^{56}\mathrm{Ca}(2\mathrm{ e}^-,2\nu_\mathrm{e})^{56}\mathrm{Ar}$.  For FRDM92 mass 
model the second electron capture
gives $6.37$\,MeV (per nucleus),
whereas for HFB24 mass model only $0.71$\,MeV is released 
(the difference in the masses of $^{56}\mathrm{K}$ and $^{56}\mathrm{Ar}$ for two models is visible in the mass charts of Fig.\ \ref{Fig_MB}). On the contrary, for the extreme rp-process ashes all applied mass models predict almost the same $Q_\mathrm{o}(P_\mathrm{oi})$ dependence. For FRDM12 and FRDM92 mass models heating profiles for superburst and KEPLER ashes become rather similar (see first two panels in Fig. \ref{Fig:Qo}). 

For all the considered mass models and $P_\mathrm{oi}\gtrsim 10^{29}$~dyn\,cm$^{-2}$, extreme rp-process ashes 
lead to
the lowest $Q_o(P_\mathrm{oi})$ among all studied ash compositions. Taking $P_{\rm oi}=P_\mathrm{nd}^\mathrm{(cat)}$, the heat release in the outer crust can be estimated as $Q_\mathrm{o}\sim(0.14-0.15)$\,MeV per accreted baryon for mixtures of elements with abundance peak at $A=106$,  and $Q_\mathrm{o}\sim (0.15-0.27)$\,MeV in the case $A\approx60$.
Quite similar values can be obtained 
by using the heat release profiles
from the one-component model
of \cite{HZ08} or \cite{Fantina_ea18} ($Q_\mathrm{o}\sim0.13$\,MeV for ashes composed of $\rm{^{106}}Pd$, and $Q_\mathrm{o}\sim (0.13-0.20)$\,MeV for $\rm{^{56}}Fe$, all per accreted baryon; see also GC21). 

At the same time, somewhat higher values are presented
by \cite{lau_ea18}: $Q_\mathrm{o}\sim0.24$\,MeV per baryon for extreme rp-process ashes 
and $Q_\mathrm{o}\sim0.29$\,MeV for superburst and KEPLER ashes 
(the values are read out from 
their figures
for the density corresponding to  $P_\mathrm{nd}^\mathrm{(cat)}$ for different ashes).
The difference is likely related to the fact, that these authors include excited states in their simulations and treated separately electron captures and neutron emission processes. However, as discussed in \cite{SC19_MNRAS}, for some transitions (for example, for beta-captures by $^{56}$Fe) the theoretical model applied in \cite{lau_ea18} overestimates the energy of the first excited state, leading to unrealistic increase of the heating associated with this reaction. We checked that for superburst ashes one of the sources of the discrepancy between our results and those of \cite{lau_ea18} is the delayed beta-captures by $\rm{^{56}}Fe$ in \cite{lau_ea18}.
We also
checked that, generally, the composition profiles and the reaction network
pathways predicted by our simplified reaction network
for FRDM92 mass model are in a reasonable agreement with
the results shown in  \cite{lau_ea18},
which are based on the same mass model.
In particular, the impurity parameter profile agrees very well with that calculated by  \cite{lau_ea18} (see  Fig.~\ref{Fig_Lau} in Appendix \ref{app_compos}).
This confirms the applicability of our model as a tool for simplified modelling of the reaction network in the crust.

In Appendix \ref{app_compos} we present 
results of our detailed calculations 
of the  average charge and impurity parameter profiles for all the considered models. These data are required for modelling the thermal evolution of NS crust.
It should be noted that the average charge profile only weakly depends on the theoretical mass model (see Fig. \ref{Fig_Z}). In contrast, the impurity parameter
appears to be rather sensitive to the employed mass model
at sufficently large pressures near the outer/inner crust interface
(Fig.\ \ref{Fig_Qimp}).

\subsection{Deep crustal heating in the nHD approach}\label{Sec:Qi}

\begin{figure}
	\includegraphics[width=\columnwidth]{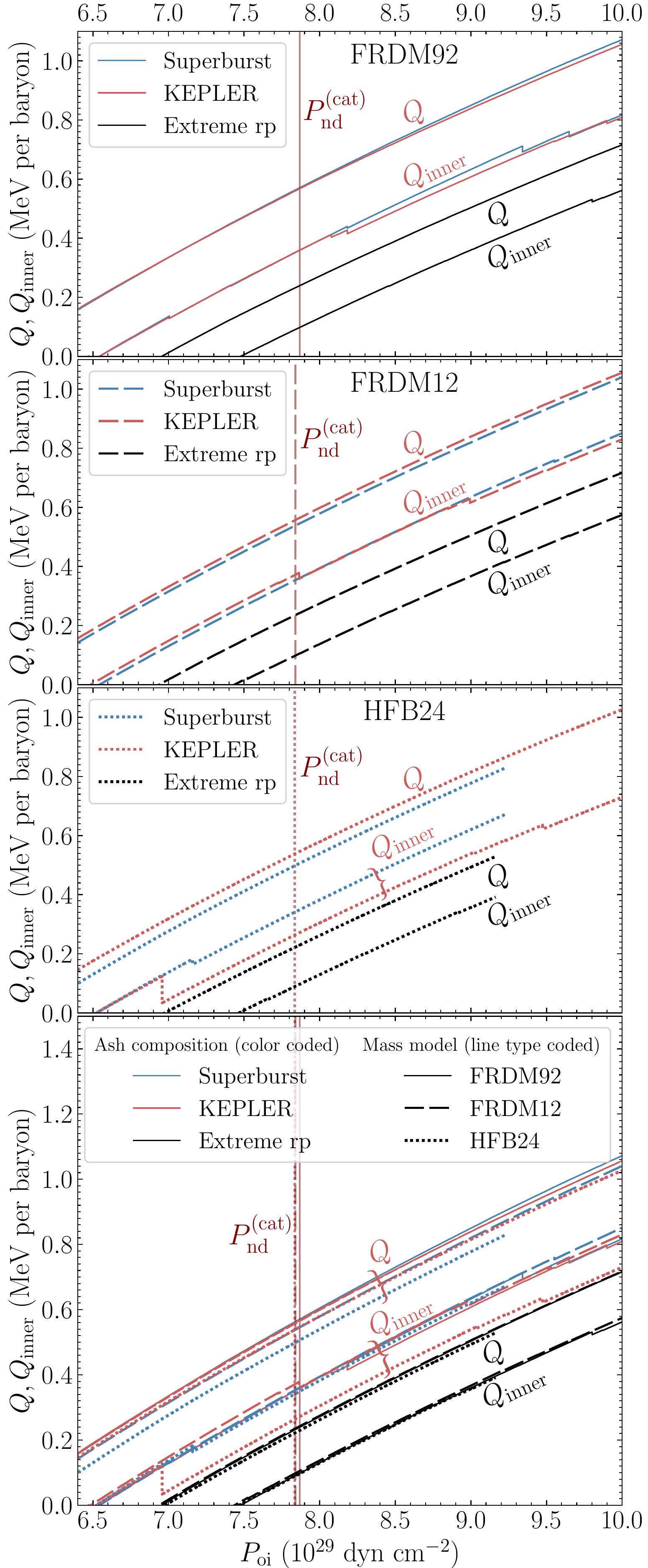}
	\caption{$Q$ and $Q_\mathrm {inner}$ vs outer-inner crust transition pressure $P_\mathrm{oi}$.  
		As in Fig.\ \ref{Fig:Qo}, 
		the three upper panels represent results for the respective mass model, 
		while the bottom panel combines all the results in one plot. 
		Ash compositions are colour coded 
		(superburst -- blue; KEPLER -- red; extreme rp -- black);
		the mass models are line type coded 
		(FRDM92 -- solid; FRDM12 -- dashed; HFB24 -- dotted lines). 
		Vertical lines indicate pressure at the outer-inner crust interface for the cold catalysed crust (line type corresponds to the mass model).}
	\label{Fig:Qi}
\end{figure} 

Let us start with the accurate definition of the total deep crustal heat release $Q$
following GC21. 
GC21 introduces the redshifted heat release $Q^\infty$ (as seen by a distant observer) 
which, strictly speaking, depends on the NS mass and EOS in the core.
However, for nHD models this dependence can be removed 
if we define the `local' quantity $Q$
according to the formula: $Q=Q^\infty e^{-\nu_\mathrm{oi}/2}$, where $e^{\nu_\mathrm{oi}/2}$ is the redshift factor at the oi interface (see equation (4) in GC21). 
The factor $e^{\nu_\mathrm{oi}/2}$ almost coincides with the redshift factor at the surface. Assuming, for instance, an NS with BSk24 EOS in the core and crust, the latter varies from 0.86 to 0.76 for NSs in the mass range $1.1 M_{\sun} - 1.8 M_{\sun}$.
The parameter $Q$ will be referred to as the total deep crustal heat release in what follows.
As shown in GC21, in the regime of a FA crust $Q$ can be parametrized by the pressure $P_\mathrm{oi}$.

According to GC21, $Q^\infty$ 
can be presented as a sum of 
three 
redshifted heat releases: 
(a) in the outer crust, $Q^\infty_\mathrm{o}\approx Q_\mathrm{o}\,e^{\nu_\mathrm{oi}/2}$
(accurate calculation of $Q^\infty_\mathrm{o}$ should take into account variation of the redshift factor in the outer crust, but it leads to negligible effect, see section II in the supplementary material of GC21);
(b) $Q^\infty_\mathrm{oi}$ at the oi interface; and (c) $Q^\infty_\mathrm{i}$ in the inner crust (including the heat released as a result of the instability leading to disintegration of 
nuclei, see \citealt{GC20_DiffEq}; GC21). Let us denote
$Q^\infty_\mathrm{inner}=Q^\infty_\mathrm{oi}+Q^\infty_\mathrm{i}$
and apply a simple expression for this quantity, derived in section III of the supplementary material in GC21:%
\begin{equation}
\label{Qi}
Q^\infty_\mathrm{inner}= e^{\nu_\mathrm{oi}/2}\left[\mu_\mathrm{b}(P_\mathrm{oi})-m_\mathrm{n}\right].
\end{equation}      
Here $\mu_\mathrm{b}(P_\mathrm{oi}) $ is the baryon chemical potential at the bottom of the outer crust and
$m_{\rm n} $ is the neutron mass.
Equation (\ref{Qi}) can be interpreted as a heat release in an abstract `reaction' that summarizes all the reactions at the outer-inner crust interface and in the inner crust, and eventually converts all the upcoming nuclei (with the redshifted baryon chemical potential  $\mu_\mathrm{b}(P_\mathrm{oi}) e^{\nu_\mathrm{oi}/2}$) into baryons in the core (thanks to the nHD condition,  the redshifted baryon chemical potential in the core equals $m_\mathrm{n} e^{\nu_\mathrm{oi}/2}$).

The composition at the bottom of the outer crust is calculated in Section \ref{Sec:outercrust}, making $\mu_\mathrm b$ a known function of $P_\mathrm{oi}$ and thus allowing us to apply Eq.\ (\ref{Qi}). The results are shown in Fig.\ \ref{Fig:Qi}, where we present 
$Q_\mathrm {inner}=Q^\infty_\mathrm {inner}\, e^{-\nu_\mathrm{oi}/2}$ and $Q=Q_\mathrm o+Q_\mathrm {inner}$ as functions of $P_\mathrm{oi}$ for all the considered mass models and ash compositions. In the bottom panel, which combines all the results in one plot, the groups of lines, corresponding to $Q_\mathrm  {inner}$ and $Q$ are marked by braces. For HFB24 mass model dotted lines are ended at $P=P^\mathrm{(acc)}_\mathrm{nd}$; for other mass models $P^\mathrm{(acc)}_\mathrm{nd}$ lies outside the plot.

One sees that all ashes lead to approximately linear $Q_\mathrm{inner}(P_\mathrm{oi})$ dependence. The only exception is KEPLER ashes in HFB24 model, for which $Q_\mathrm{inner}$ drops by $\approx 0.09$~MeV per baryon at $P\approx 7\times 10^{29}$~dyn\,cm$^{-2}$. 
The drop is associated with the electron captures 
with simultaneous emission of neutrons for $\rm{^{32}}Ne$
at such $P$. This process converts $\rm{^{32}}Ne$ into $\rm{^{28}}O$, which undergoes pycnonuclear fusion, leading to formation of $\rm{^{52}}S$ and free neutrons (the neutrons are finally absorbed by other nuclei 
in the same layer). For FRDM92 and FRDM12 this process is not energetically favourable. 

Note that, in contrast to the function $Q_{\rm inner}(P_{\rm oi})$, 
the $Q$-line in Fig.\ \ref{Fig:Qi} 
always stays continuous.
It may seem strange at first glance, 
but 
it is a typical feature: 
drops of $Q_\mathrm{inner}(P_\mathrm{oi})$ are associated with localized reactions in the outer crust,
which release energy and lead to jumps of $Q_o(P_\mathrm{oi})$. As a result, the total energy release $Q$ 
is a smooth function of $P_\mathrm{oi}$. 
The continuity of $Q(P_\mathrm{oi})$ also follows directly from the equation (4) of GC21.

For the extreme rp-process ashes, $Q_\mathrm{inner}$ weakly depends on the nuclear mass model in the outer crust, being smaller than for the superburst and KEPLER ashes by $\sim0.3$~MeV per baryon for the same $P_\mathrm{oi}$. This difference is related to the higher mass number of nuclei in extreme rp-process ashes, which makes the corresponding nuclear composition at the bottom of the outer crust closer to the ground state. In particular, the ground state element at the outer-inner crust interface for FRDM92 and FRDM12 models is $^{118}$Kr ($Z=36$), while for HFB24 model it is $^{124}$Sr ($Z=38$; note that for other HFB models the mass number can be a bit different, see \citealt{Chamel_etal15_Drip}). 
In turn, for extreme rp-process ashes the most abundant nuclide at $P_\mathrm{oi}=P_\mathrm{nd}^\mathrm{(cat)}$ is $\rm{^{104}}Ge$ ($Z=32$) for all the considered mass models. This is in sharp contrast with the superburst and KEPLER ashes, for which the most abundant elements at the oi-interface have $Z\approx18-20$ and $A\approx 60$ (see Fig. \ref{Fig_Z}). Very different nuclear compositions at the oi-interface predicted by different ash models should strongly affect the overall composition of the inner crust, which is crucial for the determination of $P_\mathrm{oi}$ in the regime of FA crust (see GC21). We come to conclusion that the pressure $P_\mathrm{oi}$ will likely depend on the composition of ashes.

Similarly to GC21, we set a lower bound on the pressure $P_\mathrm{oi}$ by requiring $Q_\mathrm{inner}>0$. It gives $P_\mathrm{oi}>6.5\times 10^{29}$~dyn\,cm$^{-2}$
for superburst and KEPLER ashes and $P_\mathrm{oi}>7.4\times 10^{29}$~dyn\,cm$^{-2}$ for extreme rp-process ash. 
According to Fig. \ref{Fig:Qi}, these constraints correspond to $Q > 0.13-0.2$~MeV per baryon. This lower bound is an order of magnitude smaller than in the traditional approach and, probably, is not very restrictive (too small). We expect that a much more interesting lower bound on $P_\mathrm{oi}$ will be obtained by calculating the heat release at the oi-interface and in the shallow layer of the inner crust (where the theoretical atomic mass tables still can be trusted), and requiring that the remaining heat released in the deeper layers of the inner crust to be non-negative. We plan to perform such an analysis in our subsequent publication.

In GC21 we suggest that the catalyzed pressure $P_\mathrm{nd}^\mathrm{(cat)}$ can serve as a conservative estimate for the upper bound on $P_\mathrm{oi}$. This conclusion is based on the results of detailed calculations of the inner crust composition performed within the CLD+sh model of \cite{carreau_ea20_Cryst_CLDsh}  for pure $^{56}$Fe nuclear ashes.  Strictly speaking, applicability of this bound for other ash compositions should be checked, especially for extreme rp-process ash enriched by heavy nuclei. Anyhow, the respective values of the heating corresponding to $P_\mathrm{oi}=P_\mathrm{nd}^\mathrm{(cat)}$ can be easily read out from Fig.\ \ref{Fig:Qi}: $Q\sim0.54$~MeV per baryon for superburst and KEPLER ashes
and $Q \sim0.24$~MeV per baryon for extreme rp-process ashes
(cf.\ the value $1.5-2.0$ MeV per baryon obtained within the traditional approach).


\section{Summary}

We apply the multi-component simplified reaction network 
to find the heat release $Q_\mathrm{o}$ 
and composition 
(in particular, the impurity parameter, $Q_{\rm imp}$,
and average charge, $\langle Z \rangle$)
in the outer crust of accreting NS 
for different compositions of thermonuclear burning ashes 
(superburst, KEPLER, and extreme rp-process ashes). 
Calculations are made for the three theoretical atomic mass tables 
(FRDM92, FRDM12, and HFB24) 
to check the sensitivity of the results to the chosen nuclear mass model 
(note, however, that for nuclei with experimentally 
measured
masses AME20 is applied).

We find, first of all, that the average charge profile in the outer crust is rather insensitive to the applied mass model (Fig.\ \ref{Fig_Z}). This is in contrast to the strong model dependence of the heat release (Figs.\ \ref{Fig:Qo}, \ref{Fig_AME16_Q}) and impurity parameter profiles (Fig.\ \ref{Fig_Qimp}). 
It is notable that the latter parameter, obtained in our simplified reaction network, agrees well with $Q_{\rm imp}$
calculated by \cite{lau_ea18} if the same mass model (FRDM92) is applied (Fig.\ \ref{Fig_Lau}).
The fact that the heat release $Q_{\rm o}$ and impurity parameter $Q_{\rm imp}$ are sensitive to the employed theoretical mass model suggests that, 
currently,
an uncertainty in determination of these quantities in the outer crust is rather dominated by the uncertainties in the mass models, than by the details of the reaction kinetics.

Having at hand 
$Q_{\rm o}$,
we
determine the total deep crustal heat release, $Q$, and the heat release
in the inner crust (including the outer-inner crust interface contribution), 
$Q_\mathrm{inner}$, in the regime of a fully accreted crust following the thermodynamically consistent approach developed in GC21.

Our results for $Q$, $Q_\mathrm{o}$ and $Q_\mathrm{inner}$ are parametrized by the pressure $P_\mathrm{oi}$ at the outer-inner crust interface 
(Figs.\ \ref{Fig:Qo}, \ref{Fig:Qi}).
As argued in GC21, accurate calculation of $P_\mathrm{oi}$ depends sensitively on 
the inner crust EOS, which  
has not yet been analysed within the nHD approach for complex ash compositions.
In this work we, therefore, treat $P_\mathrm{oi}$ as a free parameter. We constrain $P_\mathrm{oi}$  from below by the requirement $Q_\mathrm{inner}>0$, which gives: $P_\mathrm{oi}>6.5\times 10^{29}$~dyn\,cm$^{-2}$ for superburst and KEPLER ashes and $P_\mathrm{oi}>7.4\times 10^{29}$~dyn\,cm$^{-2}$ for extreme rp-process ashes, implying  $Q>0.13-0.2$~MeV per baryon. It is much more complicated to constrain $P_\mathrm{oi}$ from above (see Section \ref{Sec:Qi}). 

For the same $P_\mathrm{oi}$ superburst and KEPLER ashes lead to a larger heating than the extreme rp-process ash (the difference is $\sim 0.3$~MeV per baryon);
however, we warn the reader, that the actual $P_\mathrm{oi}$ likely depends on the ash composition and this can (partially) compensate the difference 
(for instance, such partial compensation takes place for the lower bounds on $Q$ quoted above).

An important advantage of our approach, based on GC21, is that it encodes all uncertain physics of deep inner crust layers in
just one parameter, $P_\mathrm{oi}$. 
In our calculations, we deal
mostly 
with the masses
of neutron-rich isotopes
before the neutron-drip line. 
There is great progress in experimental measurements of the masses of such nuclei 
(e.g., \citealt{Meisel_ea20_mass_measurement}) and new experiments are planned (\citealt{Kim20_RAON,Meisel20_FRIB}). 
This gives us hope that the profiles $Q(P_\mathrm{oi})$ and $Q_\mathrm{inner}(P_\mathrm{oi})$ will become more certain in the not-too-distant future (even nowadays the uncertainty associated with the theoretical mass models is
rather modest, see Fig.\ \ref{Fig:Qi}). This opens up an attractive possibility to constrain $P_\mathrm{oi}$ by comparing observations of crustal coolers with the theoretical modelling of their thermal evolution.

In a subsequent publication, we plan to continue and extend this work by analysing the shallow layers of the inner crust and calculating the heating profile and nuclear composition there. Apart from the fact that this information is vitally important for modelling 
the thermal evolution of transiently accreting NSs, 
we expect that it will also allow us to tighten a lower bound 
on $P_\mathrm{oi}$ by imposing the condition that the heat release in the remaining part of the inner crust should be positive.

\section*{Acknowledgements}
The work of N.~N.\ Shchechilin was supported by the Foundation for the Advancement of Theoretical Physics and Mathematics ``BASIS'' (grant \#20-1-5-79-1).
The work of M.~E.\ Gusakov was supported by RFBR [Grant No.\ 19-52-12013].
\section*{DATA AVAILABILITY}
Data can be provided by the authors upon a reasonable request.

\bibliographystyle{mnras}

\appendix
\newpage

\section{
Impurity parameter and average charge}\label{app_compos}

\begin{figure}
	\includegraphics[width=\columnwidth]{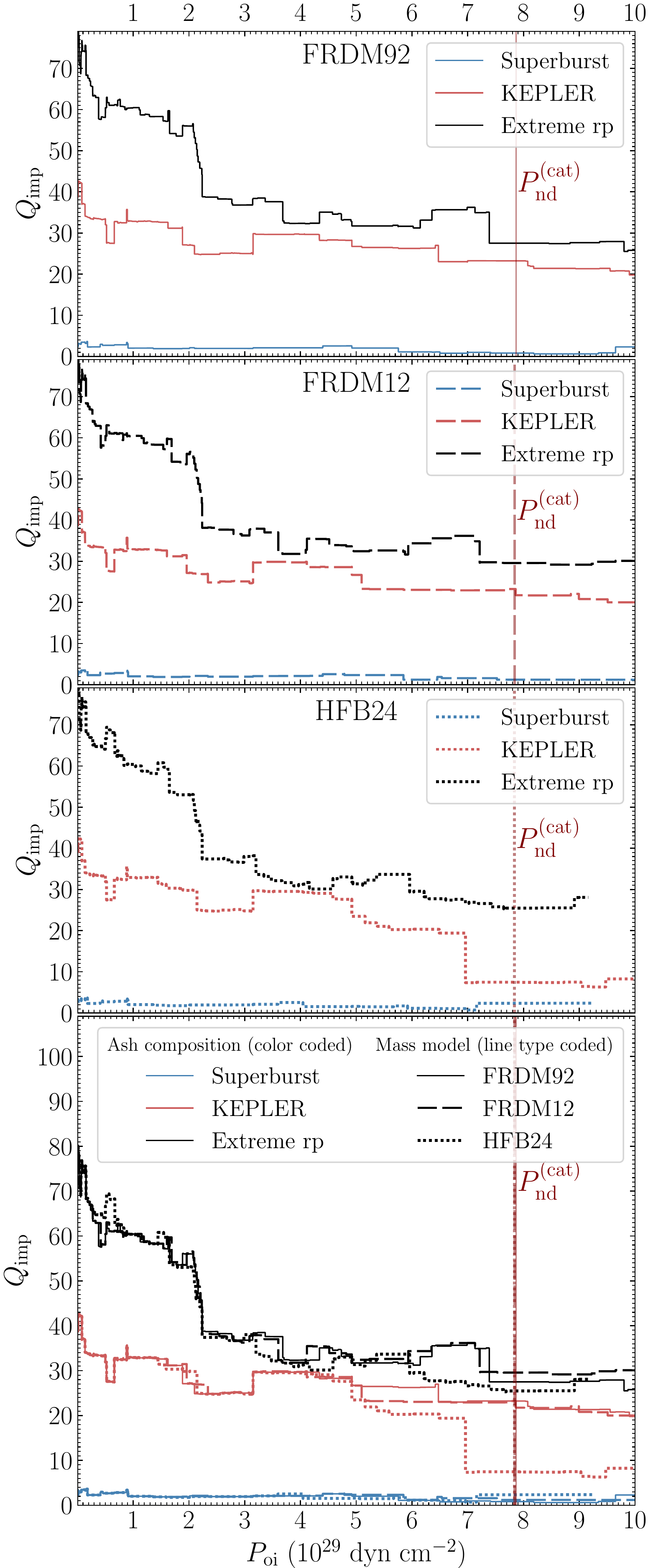}
	\caption{$Q_\mathrm{imp}$ vs $P_\mathrm{oi}$ for the three ash compositions considered in the paper.  
		As in Fig.\ \ref{Fig:Qo}, 
		the three upper panels represent results for the respective mass model, while the bottom panel combines all the results in one plot. 
		Ash compositions are colour coded 
		(superburst -- blue; KEPLER -- red; extreme rp -- black);
		the mass models are line type coded 
		(FRDM92 -- solid; FRDM12 -- dashed; HFB24 -- dotted lines). 
		Vertical lines indicate pressure at the outer-inner crust interface for the cold catalysed crust (line type corresponds to the mass model).}
	\label{Fig_Qimp}
\end{figure}

\begin{figure}
	\includegraphics[width=\columnwidth]{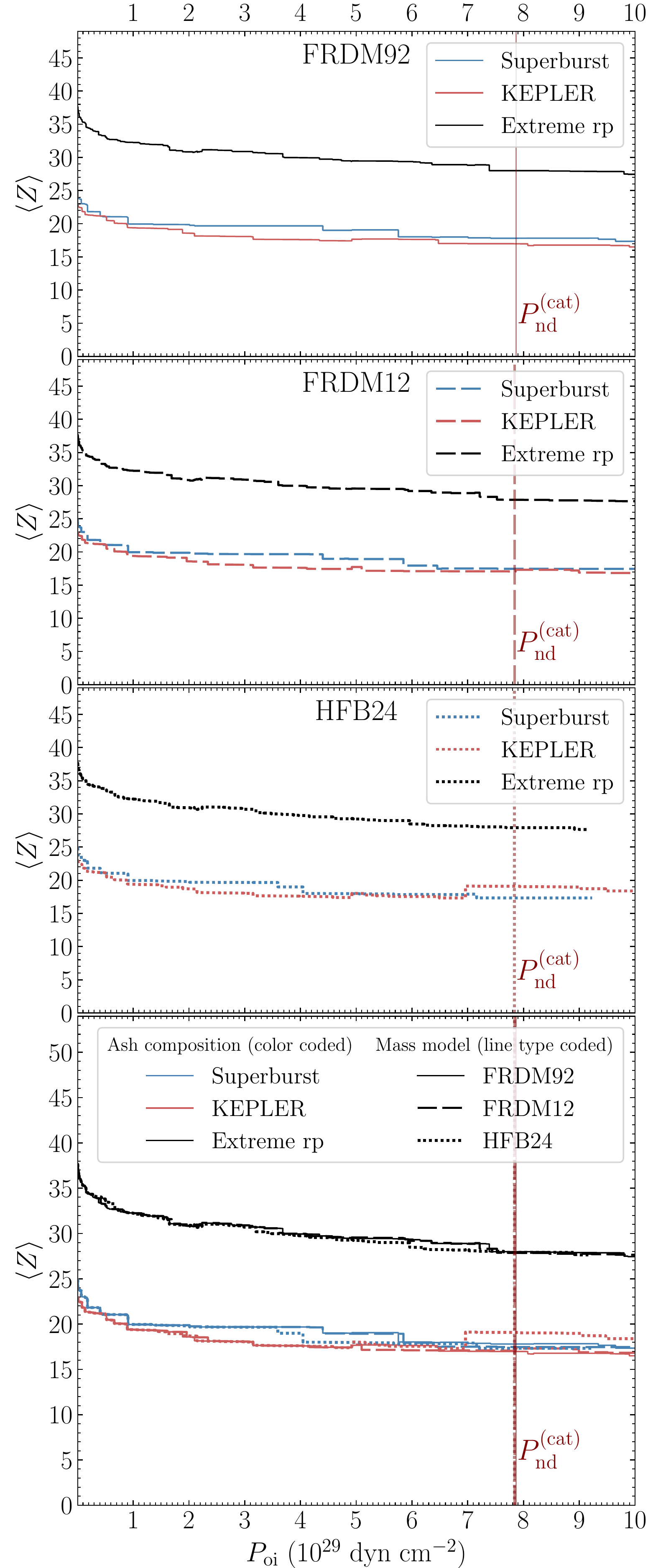}
	\caption{$\langle Z\rangle$ vs $P_\mathrm{oi}$ for the three ash compositions considered in the paper. The notations are the same as in Fig.\ \ref{Fig_Qimp}. }
	\label{Fig_Z}
\end{figure}

For completeness, in Figs.\ \ref{Fig_Qimp} and \ref{Fig_Z} we present profiles of the average charge $\langle Z\rangle=\sum_i X_i Z_i$ and the impurity parameter $Q_\mathrm{imp}=\sum_i X_i(Z_i-\langle Z\rangle)^2$, where $X_i$ denotes the fractional number of ions of type $i$.
These data can be 
useful
for modelling the thermal evolution of accreting NSs.

It is interesting that the function $Q_{\rm imp}(P_{\rm oi})$ is sensitive to the employed mass model for sufficiently large pressures ($P_{\rm oi}\gtrsim 6\times 10^{29}$~dyn~cm$^{-2}$; see bottom panel in Fig.\ \ref{Fig_Qimp}).
At the same time, $\langle Z\rangle$-profile
remains roughly 
the same for different mass models (see bottom panel in Fig.\ \ref{Fig_Z}).

\begin{figure}
	\includegraphics[width=\columnwidth]{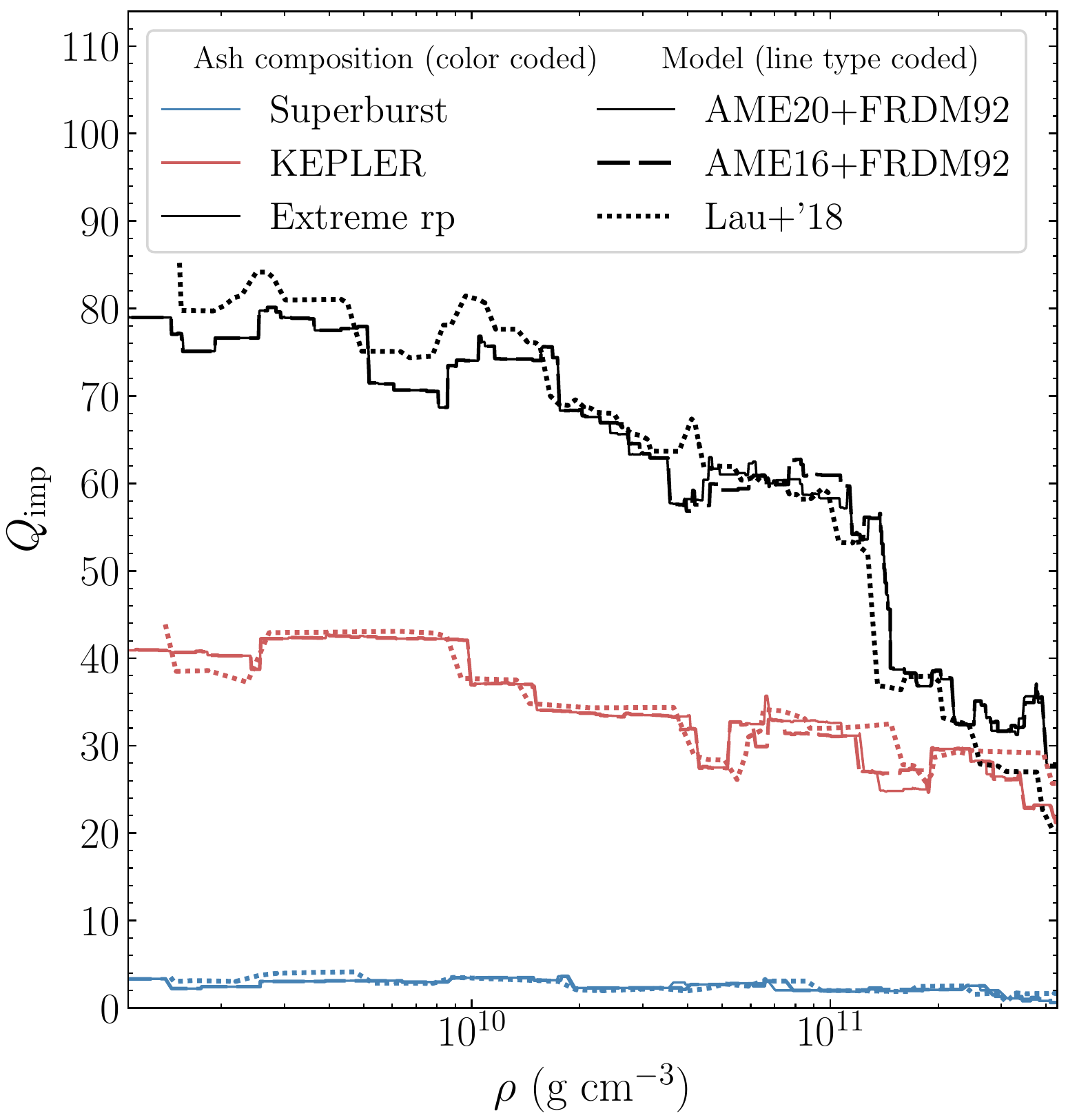}
	\caption{
	Comparison of the impurity parameter profile $Q_{\rm imp}(\rho)$ in the outer crust calculated within the simplified reaction network (solid and dashed lines), with the results by \protect\cite{lau_ea18} (dotted lines).
	Ash compositions are color coded (superburst -- blue; KEPLER -- red; extreme rp -- black); the models are line type coded 
	(AME20+FRDM92 -- solid; AME16+FRDM92 -- dashed; Lau+'18 -- dotted lines).}
	\label{Fig_Lau}
\end{figure}

In Fig.\ \ref{Fig_Lau} we compare profiles of $Q_{\rm imp}$ calculated within our reaction network (solid and dashed lines) with $Q_{\rm imp}$ presented by \cite{lau_ea18}
(dots; the corresponding data were read out from their figure 24).
To plot Fig.\ \ref{Fig_Lau}, we adopt 
the same 
theoretical mass table FRDM92 as was used by
\cite{lau_ea18}, but supplement it with
AME16 (dashes) or AME20 (solid lines) 
experimental mass tables.
One sees a very good agreement between all the three calculations.
This means that the simplified reaction network developed in this work can be used for reliable calculations of the impurity parameter and average charge profiles in the outer crust.


\newpage

\section{Comparison of the employed theoretical mass models and importance of the progress in experimental nuclear mass measurements}\label{app_massModels}
\begin{figure*}
	\includegraphics[width=0.9\textwidth]{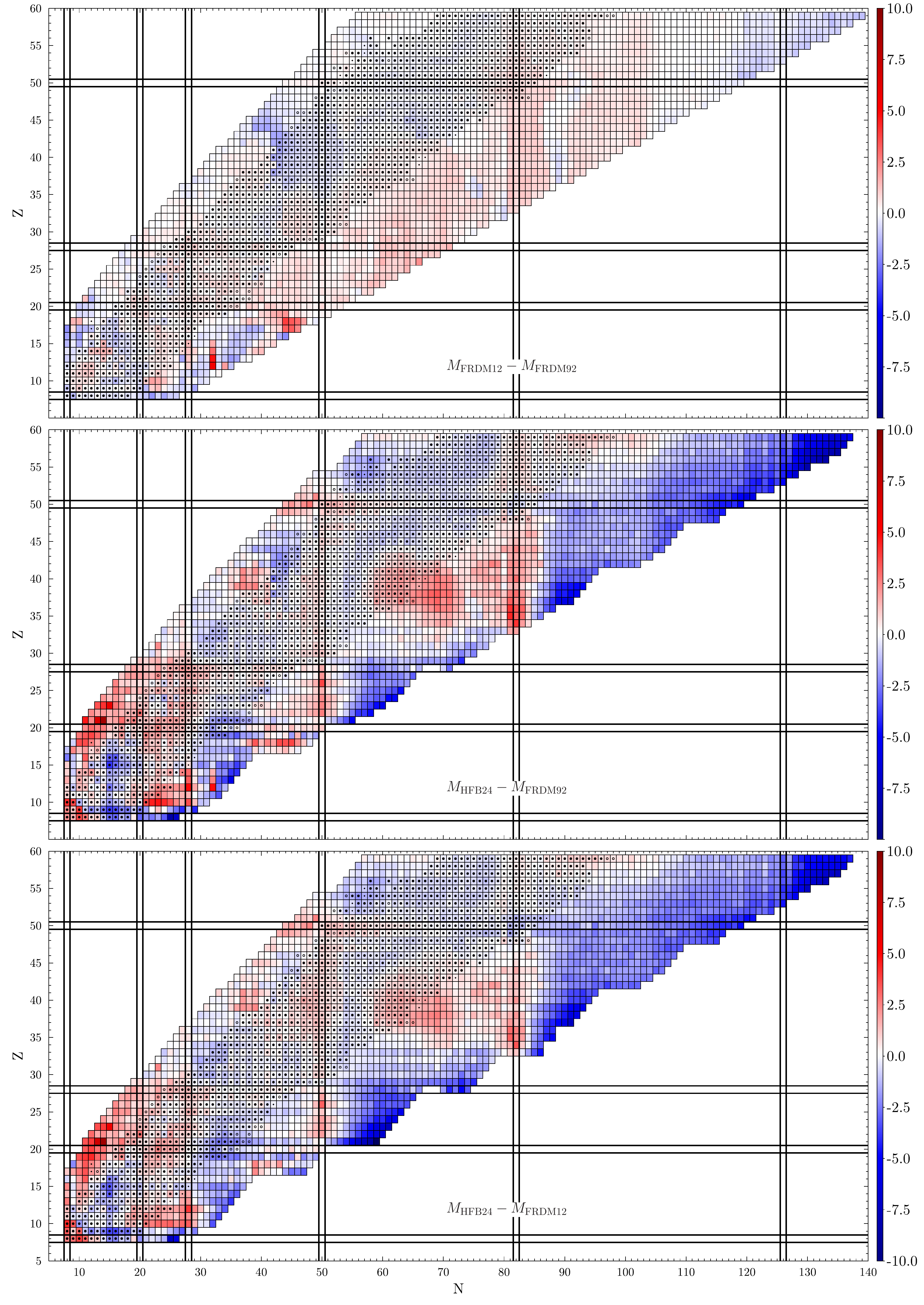}
	\caption{Mass differences (in MeV)  
		for theoretical mass models employed in this work.
		Magic numbers are shown by pairs of thick solid lines.
		Nuclei with known experimental masses are indicated by symbols in the centre of respective square. Open circles, small dots, and filled circles are for nuclei that are presented only in AME20, only in AME16, 
		and in both AME20 and AME16 tables, respectively.}
	\label{Fig_MB}
\end{figure*}

Figure \ref{Fig_MB} compares the theoretical mass models employed in this work.
One can notice substantial uncertainties in the mass values of the neutron-rich nuclei, used in our calculations.
This uncertainty increases with increase of the neutron  excess and distance from nuclei with experimentally measured masses (indicated by open circles, small dots and filled circles; see caption to the figure).

\begin{figure}
	\includegraphics[width=\columnwidth]{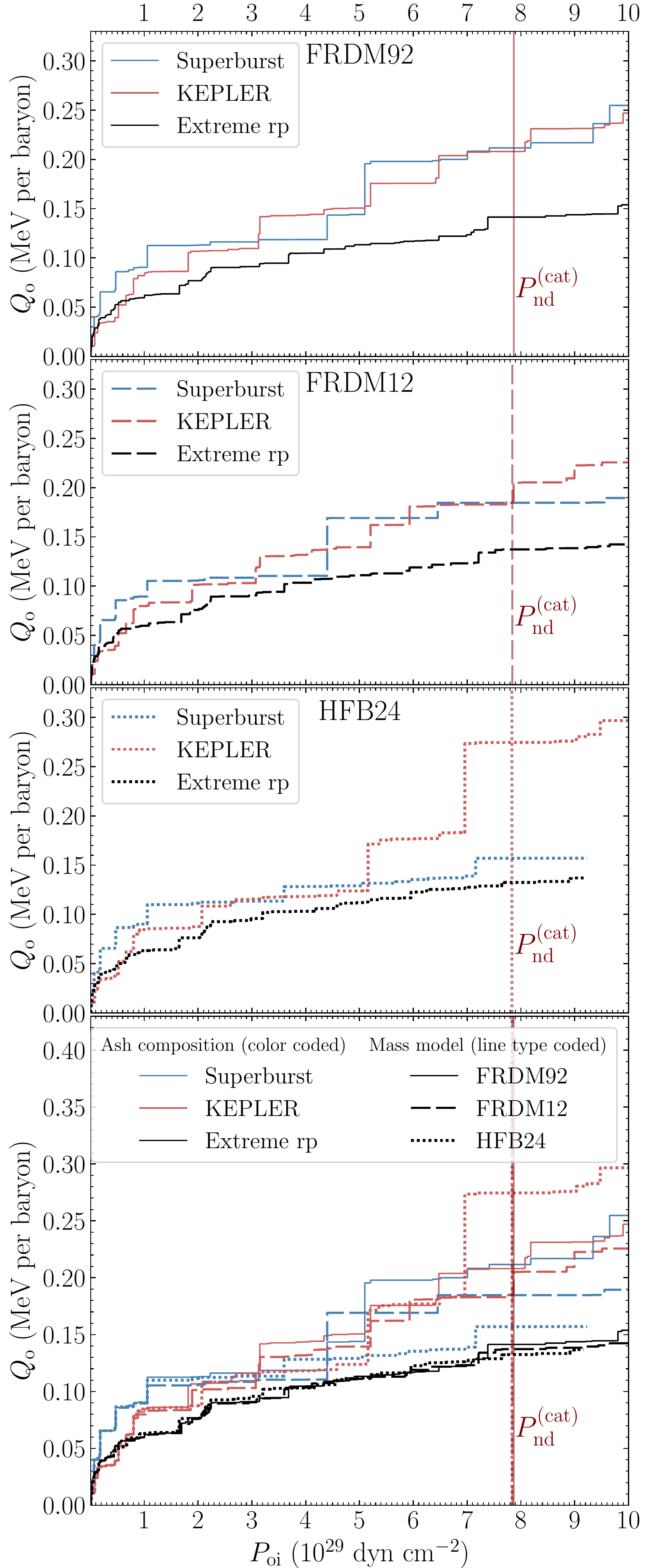}
	\caption{Analogue of the  figure \ref{Fig:Qo}, but with AME16 applied as the experimental atomic mass table.}
	\label{Fig_AME16_Q}
\end{figure}

\begin{figure}
	\includegraphics[width=\columnwidth]{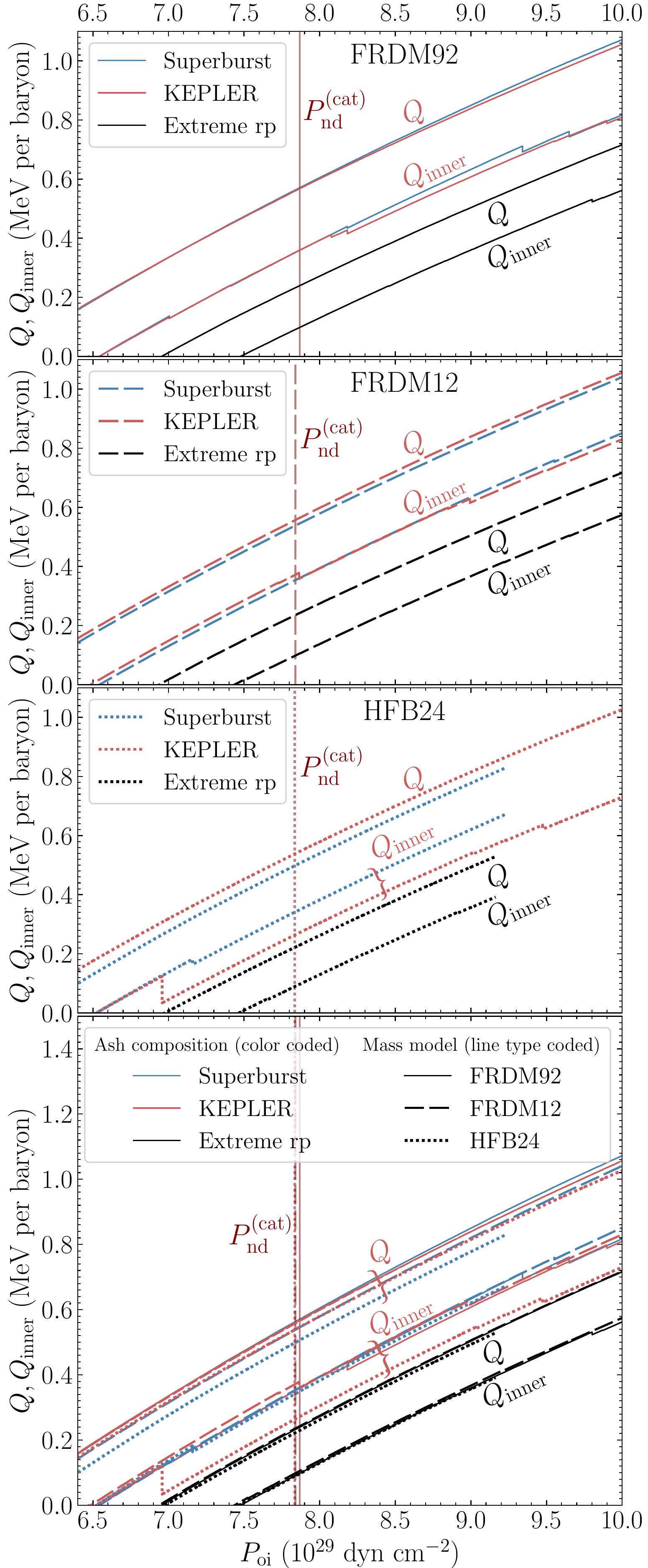}
	\caption{Analogue of the figure \ref{Fig:Qi}, but with AME16 applied as the experimental atomic mass table.}
	\label{Fig_AME16_mu}
\end{figure}

Recently, the experimental mass table AME16 has been replaced by the new one, AME20. To demonstrate importance of this update, 
below 
we present analogues of Figs.\ \ref{Fig:Qo} and \ref{Fig:Qi}, calculated 
using
AME16 instead of AME20 (see Figs.\ \ref{Fig_AME16_Q} and \ref{Fig_AME16_mu}).  
We confronted these figures and found noticeable changes in the energy release profile and outer crust composition for AME20 in comparison to AME16 table (however, qualitatively, the results remain unchanged).
In particular, the difference between the curves in the figures 
corresponding to different theoretical mass models is slightly reduced for AME20,
emphasizing the role of new experimental nuclear mass measurements for astrophysical applications.

\label{lastpage}
\end{document}